# One-Dimensional Ionic-Bonded Structures in NiSe Nanowire

Xiaozhi Liu,[1,a)] Ang Gao,[2,a)] Qinghua Zhang,[1] Yaxian Wang,[1] Yangyang Zhang,[3] Yangfan Li,[1,4] Xing Zhang,[5] Lin Gu,[6] Jinsong Hu,[4,5] and Dong Su[1,4,b)]

AFFILIATIONS

[1]Beijing National Laboratory for Condensed Matter Physics, Institute of Physics, Chinese Academy of Sciences, Beijing 100190, China

[2]Institute of Environmental and Applied Chemistry, College of Chemistry, Central China Normal University, Wuhan 430079, China

[3]College of Chemistry, Henan Institute of Advanced Technology, Zhengzhou University, Zhengzhou 450052, China

[4]University of Chinese Academy of Sciences, Beijing 100049, China

[5]Beijing National Laboratory for Molecular Sciences, CAS Key Laboratory of Molecular Nanostructure and Nanotechnology, Institute of Chemistry, Chinese Academy of Sciences, Beijing 100190, China

[6]Beijing National Center for Electron Microscopy and Laboratory of Advanced Materials, School of Materials Science and Engineering, Tsinghua University, Beijing 100084, China

[a)]These authors contributed equally to this work.

[b)]Authors to whom correspondence should be addressed: dongsu@iphy.ac.cn

# ABSTRACT

One-dimensional van der Waals (1D vdW) materials, characterized by atomic chains bonded ionically or covalently in one direction and held together by van der Waals interactions in the perpendicular directions, have recently gained intensive attention due to their exceptional functions. In this work, we report the discovery of 1D ionic-bonded structures in NiSe nanowires. Utilizing aberration-corrected scanning transmission electron microscopy, we identified four distinct structural phases composed of two fundamental 1D building blocks: a triangle-shaped unit and a parallelogram-shaped unit. These phases can transform into one another through topotactic combinations of the structural units. Density functional theory calculations reveal that these structural units are bound by ionic bonds, unlike the van der Waals forces typically found in 1D vdW materials. The diverse arrangements of these building blocks may give rise to unique electronic structures and magnetic properties, paving the way for designing advanced materials with novel functionalities.

The crystal structure of materials is considered the primary factor influencing their properties in electrical, optical, catalytic, and magnetic applications.[1-3] Recently, one-dimensional (1D) van der Waals (vdW) crystals have garnered significant interest due to their ability to tune crystal structures beyond the constraints of conventional crystallography.[4, 5] 1D vdW crystals are formed by stacking electron-confined 1D building units or atomic chains.[6, 7] The spacing between these building units typically ranges from 3 to 4 Å, held together by van der Waals forces.[8] It is believed that these weak vdW interactions are crucial for tuning their structures and enabling their unique physical properties.[9] Structurally, 1D vdW crystals exhibit a highly anisotropic arrangement; however, whether such highly anisotropic structures can exist in non-vdW systems remains an open question.

In this work, we explore how the concept of a 1D crystal with a highly anisotropic structure can be extended to a system of NiSe, where ionic bonds hold the building blocks together. Using atomic-resolution analytical scanning transmission electron microscopy (STEM), we identify four distinct 1D crystal structures constructed from two types of units, referred to as NiSe-Δ and NiSe-I, respectively. Density functional theory (DFT) calculations suggest that most electrons are localized within the NiSe-Δ and NiSe-I units. While the inter-unit bonds are ionic, they are weaker than the intra-unit ionic bonds. This work uncovers new kinds of 1D building blocks and broadens our understanding that interactions between building blocks can involve not only van der Waals forces but also chemical ionic bonds.

We synthesized NiSe nanowires on Ni foam using a facile hydrothermal method [Fig. 1(a)]. The scanning electron microscopy (SEM) image in Fig. 1(b) shows their uniform size and morphology. To ensure the air stability of NiSe and enhance its functionality in electrocatalysis, the nanowires were further coated with $MoSe_2$. As shown in the supplementary material (Fig. S1), a cross-sectional sample was prepared by a focused ion beam (FIB). The STEM-electron energy-loss spectroscopy (EELS) mapping in Fig. 1(c) reveals the elemental distributions within the NiSe@$MoSe_2$ core-shell structure. The as-synthesized nanowire composites demonstrated enhanced performance in the hydrogen evolution reaction.[10] However, the atomic structure of the NiSe phase remains unresolved. As shown in Fig. 1(d), a high-angle annular dark-field (HAADF) image reveals the atomic structure of NiSe nanowire, showing multiple boundaries and some disorderly arranged regions. The corresponding fast Fourier transform (FFT) pattern in Fig. 1(e) shows the presence of multiple crystal variants, despite the fact that the X-ray diffraction (XRD) profile in Fig. 1(f) indicates that most NiSe nanowires adopt a structure akin to the R3m space group (JCPDS No. 18-0887). A diffraction profile was generated by integrating the FFT pattern from Fig. 1(e). Compared to the simulated electron diffraction (ED) profile of an R3m NiSe structure, there are diffraction peaks that deviate from the simulation results, indicating the complexity of the internal structure.

From the atomic resolution HAADF-STEM image in Fig. 2(a), we observe four different phases in the NiSe core and denote them as phases A, B, C, and D, respectively [Fig. 2(b)]. Furthermore, we find that these phases are composed of two basic structural units. In the cross-sectional view, the projections of these units appear as a triangle (denoted as NiSe-Δ) and a parallelogram (denoted as NiSe-I), respectively. As shown in Fig. 2(b), the unit cells of the four phases can be symbolized by unique combinations of these triangles and parallelograms. The atomic arrangement of Ni and Se is justified by the atomic-resolution STEM-EELS mapping in Fig. 2(c). Based on these experimental results, we build the atomic models of NiSe-Δ, NiSe-I, and their mirrored structures, as shown in Fig. 2(d). These structural units extend along the nanowire axis, forming 1D building blocks. The simulation of the HAADF images, provided in the supplementary material (Fig. S2), further validates the structure of the four phases. High-resolution transmission electron microscopy (HRTEM) and HAADF-STEM images, captured from a view perpendicular to the nanowire axis, are shown in the supplementary material (Fig. S3). The diffuse streaks in the FFT patterns indicate the varying stacking sequences of the 1D building blocks perpendicular to the axis of the nanowires.[11, 12] Attributed to the compatible features at the edges of the structural units, different phases of NiSe can be formed by periodically arranging these building blocks.

This flexibility allows for the construction of multiple phases from the NiSe-Δ and NiSe-I structural units. By adjusting the number of parallelograms in the structure and the relative position of the triangles, various possible phases can be generated, as shown in the supplementary material (Fig. S4). Some of the proposed phases have already been observed in previously reported structure models, such as R3m[13] and Pca2$_1$[14], as depicted in the supplementary material (Fig. S5).

We developed a model to illustrate the structural correlations of phases A, B, C, and D, as shown in Figure 3(a). The topological structures of the four phases can be conceptually transformed from one to another via translation and insertion of units. For example, phase A, composed of the -I-∇-I-Δ-I- structure, transits to phase D (-I-∇-Δ-I-) via removing one parallelogram from every two. When all parallelograms in phase D are removed, phase B, composed of -∇-Δ-∇-Δ- structure, is obtained. However, the relationship between phases B and C is more complex: to transform from B to C, two structural units in the unit cell of phase B shift upward by half a unit cell, while the remaining two structural units flip 180 degrees out of plane to form phase C. We performed DFT calculations to determine the formation energies for the four phases, as detailed in the supplementary material (Table. S1). The calculated formation energies for phases A to D are -3.156, -3.151, -3.136, and -3.151 eV per atom, respectively. Phases A, B, and D have very close values of formation energy, contributing to the phenomenon of multi-phase coexistence in NiSe nanowires. In contrast, phase C has a relatively higher formation energy due to the additional energy required to maintain its shifted NiSe-Δ structural unit. According to the projected areas of the four phases in Fig. 2(a), we can roughly estimate their proportions in NiSe nanowires: 42.6% for phase A, 34.7% for phase B, 6.2% for phase C, and 16.5% for phase D. The higher formation energy of phase C accounts for its relatively small proportion in the NiSe nanowires.

Given the high similarities and strong correlations among these four phases, coherent interfaces are expected.[15-17] As shown in Fig. 3(b), four regions in the cross-section of the NiSe nanowire were selected to illustrate the interfacial structures. There are six kinds of parallel interfaces among the four phases, covering combinations of A-B, A-C, A-D, B-C, B-D, and C-D. All of these interfaces were observed in HAADF-STEM images as shown in Figs. 3(c)-3(f). It is observed that the structural units tend to form coherent interfacial structures through complementary docking.[18, 19] The NiSe-Δ and NiSe-∇ structural units can easily match with each other by sharing edges. The NiSe-I structural units can insert between the -Δ-∇- and -∇-Δ- units at the interfaces by tilting to the left or right, respectively. The flexibility of the bonding structures among the structural units guarantees the formation of coherent interfaces between different phases.[20, 21] Owing to the symmetry of triangle structural units, there can also be an angle between any two phases at the interface that is a multiple of 60 degrees.[22] By analyzing the combination of structural units, we can outline all possible interfacial structures, including parallel and angled interfaces. Regardless of the viewing directions, there are only four possible interfacial structures for any two phases, as shown in the supplementary material (Fig. S6). Three of them describe the angled interfaces between two phases, while only one describes an interface between two phases that are aligned parallel to each other. Consequently, for all four phases, there can be 18 types of angled interfaces in addition to 6 types of parallel interfaces. Some examples of the angled interfaces are shown in the supplementary material (Fig. S7).

We performed DFT calculations to investigate the electronic structure of the NiSe nanowires, employing two phases (A and B) as an example to see the similarities and differences in electronic properties resulting from their distinct atomic structures. In the supplementary material (Fig. S8), the bonding orbital states within 1 eV below the Fermi level are illustrated using a color scale. For phase A, most bonding electrons are localized within the NiSe-Δ and NiSe-I structural units. For phase B, the majority of bonding electrons are confined to the NiSe-Δ structural units. Each structural unit exhibits overall electrical neutrality. Since there are few shared bonding electrons between adjacent structural units, these units are relatively independent, justifying their designation as 'structural units'.[23-25] The electron localization function (ELF) analysis was conducted to further examine the bonding characteristics of the structural units.[26, 27] As shown in Figs. 4(a)-4(b), electrons are more localized around the Se atoms, indicating a charge transfer from Ni to Se atoms.[28, 29] As the ELF values evolve, shown in the supplementary material (Fig. S9), the isosurface consistently envelops Se atoms in a nearly spherical shape, indicating a characteristic ionic bonding between Ni and Se atoms.[30, 31] Therefore, the 1D building blocks of NiSe are most likely bound by ionic bonding force, which is an unconventional phenomenon for one-dimensional structures.[32, 33] The relatively strong coupling among building blocks may lead to uncompensated spin in NiSe nanowires. It has been reported that NiSe compounds are likely to exhibit ferromagnetism,[34-36] for which we calculate the spin-resolved density of states of phases A and B, shown in Figs. 4(c)-4(d). Phase A possesses a slightly larger asymmetry at the Fermi level.[37] The contributions of Se-$p$ and Ni-$d$ states near the Fermi level also differ between the two phases. In phase A, Se atoms contribute significantly more than Ni atoms; whereas in phase B, the contributions of Se and Ni atoms are comparable. These differences in properties originate from the structural variations between phases A and B. This insight suggests that the physical properties of NiSe can be modulated by arranging the 1D building blocks in different structural configurations.

In summary, we identified multiple 1D phases with highly anisotropic structures in NiSe nanowires. These phases consist of two types of 1D ionic-bonded structures, shaped as triangles and parallelograms. Our work demonstrates that these units serve as the building blocks of NiSe, held together by ionic bonds. This unique 1D ionic-bonded structure introduces variability in the polymorphic nature of functional materials and offers a new perspective for understanding structure-property relationships. By exploring compounds with ionic-bonded structural units and tuning their electronic structures, we anticipate uncovering a rich array of physicochemical properties.

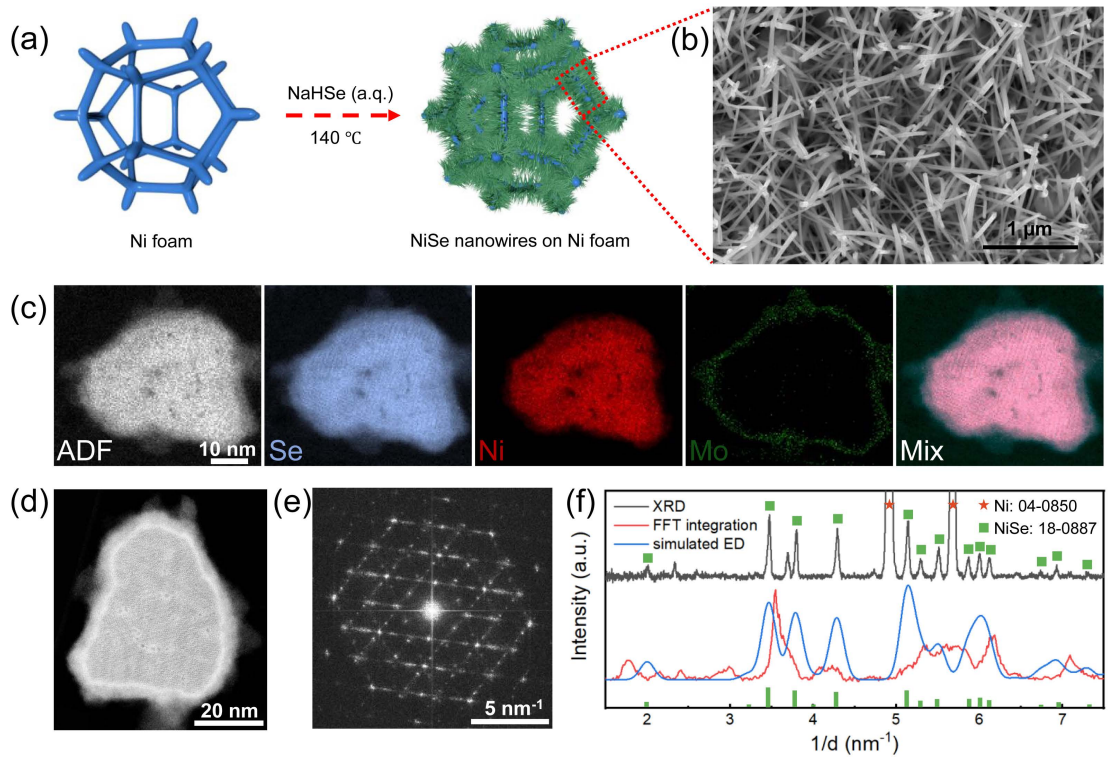

**FIG. 1.** (a) Schematic diagram of the synthesis process of NiSe nanowires. (b) SEM image of the synthesized NiSe nanowires. (c) STEM-EELS mappings of the nanowire's cross-section showing the elemental distribution of Se (grey blue), Ni (red), Mo (green), and their mixture. (d) ADF image of the NiSe nanowire and (e) their corresponding FFT pattern of (d). (f) XRD pattern of the NiSe/Ni foam (black), an integration profile of the FFT pattern in (e) (red), and a simulated electron diffraction profile of NiSe with R3m space group (blue).

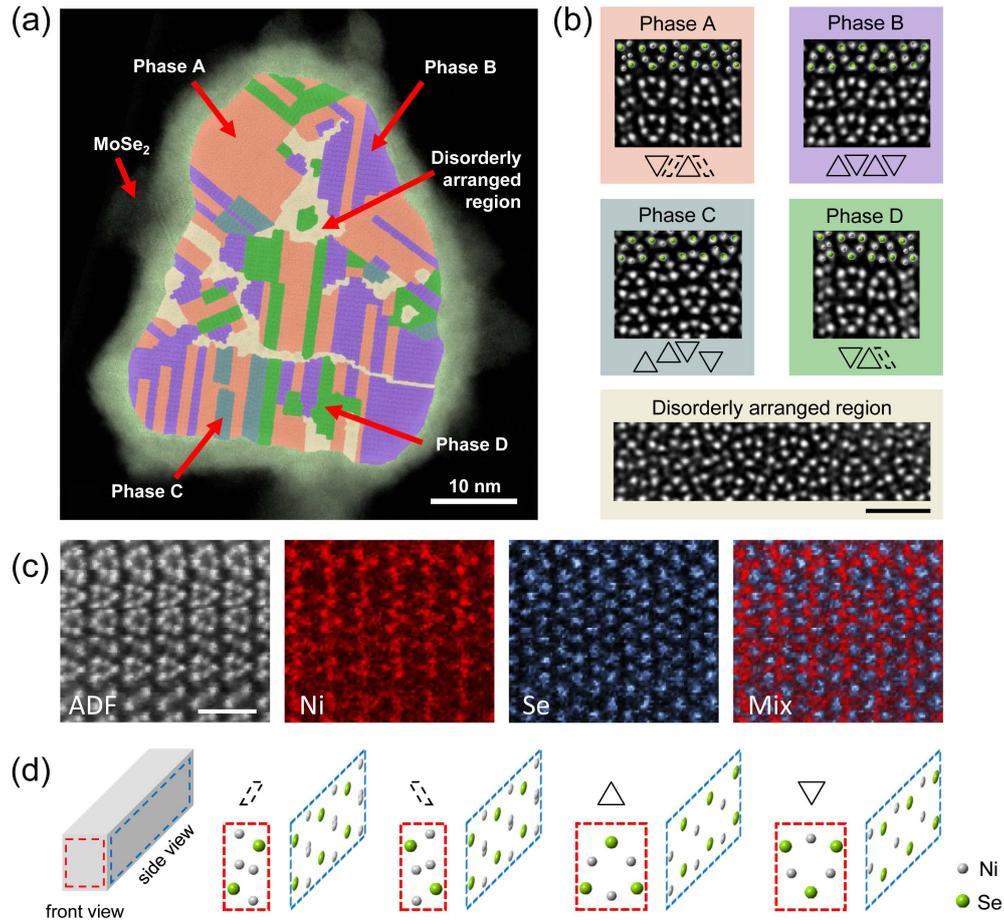

**FIG. 2.** (a) Cross-sectional HAADF-STEM image superimposed with pseudo-colors showing different phases in a NiSe nanowire. Phases A, B, C, and D are colored in red, lavender, cyan, and green, respectively. (b) Enlarged high-magnification HAADF-STEM images with superimposed atomic models. The unit cells of four phases are represented by the symbols below. The scale bar is 1 nm. (c) STEM-EELS mappings of the cross-section of the NiSe nanowire, showing the atomic resolution distributions of Ni and Se. The scale bar is 1 nm. (d) The front and side views of the atomic models for the structural units, denoted by red and blue dashed boxes, respectively. Four structural units are denoted with the right-leaning parallelogram, the left-leaning parallelogram, the upward triangle, and the downward triangle, respectively.

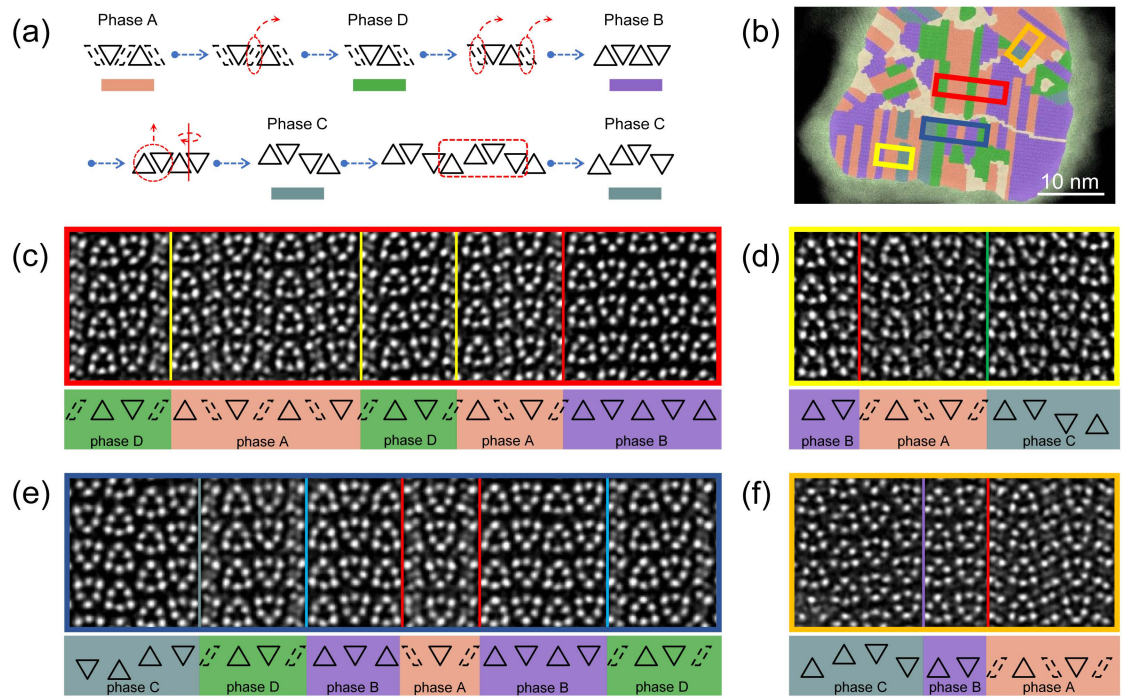

**FIG. 3.** (a) The structural correlations of the phases A, B, C, and D. (b) Selected HAADF-STEM image of the NiSe nanowire showing various coherent interfaces. (c-f) Coherent interfacial structures from different regions in (b), as indicated by the red, yellow, dark blue, and gold boxes, correspondingly. The structural units of each column are represented by symbols below. Four phases composed of these structural units are distinguished by different colors.

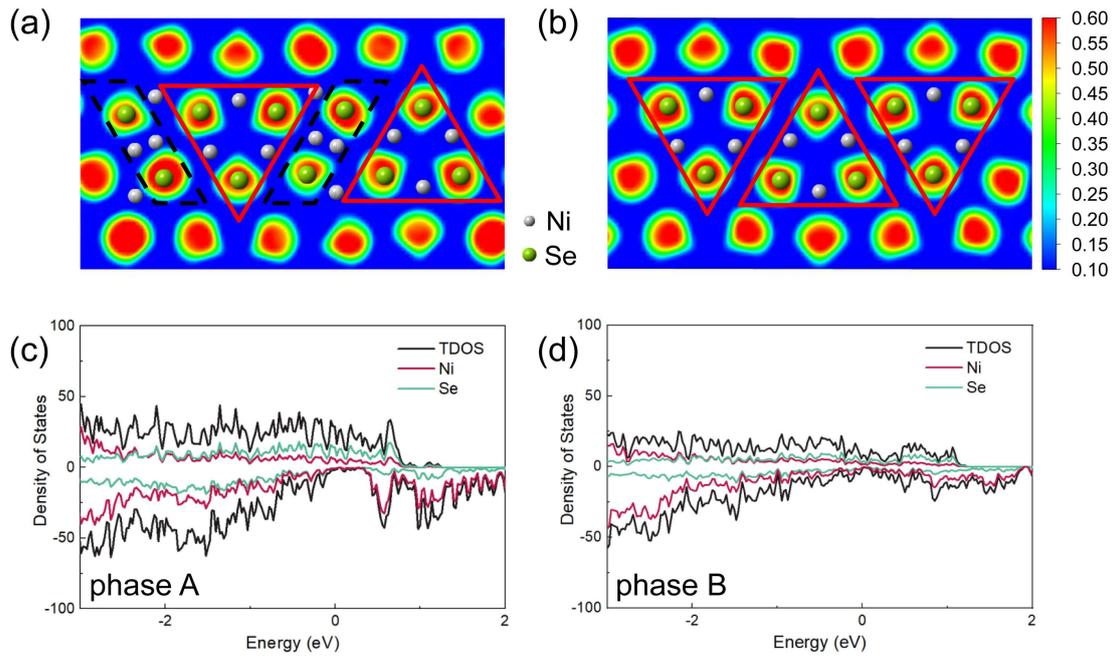

**FIG. 4.** The cross-sectional color maps of the electron localization function (ELF) of (a) phase A and (b) phase B, highlighting a strong localization on the Se atoms. The ELF values (0.1 to 0.6) are mapped on a red-green-blue color scale shown on the right. The total and atom-projected density of states of (c) phase A and (d) phase B.

## SUPPLEMENTARY MATERIAL

See the supplementary material for more details on the experimental methods, the deduction of structural models, the analysis of the angled interfaces, and visual representations of electron distribution and localization in materials.

## ACKNOWLEDGMENTS

This work was supported by the National Key Research and Development Program of China (No. 2023YFB4006203), the National Natural Science Foundation of China (No. U21A20328, 52101277, 22105220, and 22209202), and the Strategic Priority Research Program (B) (No. XDB33030200) of Chinese Academy of Sciences.

## AUTHOR DECLARATIONS

### Conflict of Interest

The authors have no conflicts to disclose.

### Author Contributions

**Xiaozhi Liu:** Investigation (lead); Methodology (equal); Visualization (equal); Writing – original draft (lead). **Ang Gao:** Formal analysis (equal); Visualization (equal); Methodology (equal). **Qinghua Zhang:** Formal analysis (equal); Investigation (equal); Methodology (equal). **Yaxian Wang:** Formal analysis (equal); Writing – review & editing (equal). **Yangyang Zhang:** Formal analysis (equal); Methodology (equal). **Yangfan Li:** Methodology (equal). **Xing Zhang:** Formal analysis (equal); Methodology (equal). **Lin Gu:** Conceptualization (equal); Supervision (equal). **Jinsong Hu:** Conceptualization (equal); Supervision (equal); Project administration (equal). **Dong Su:** Formal analysis (lead); Conceptualization (lead); Funding acquisition (lead); Supervision (lead); Project administration (lead); Resources (lead); Writing – review & editing (lead).

## DATA AVAILABILITY

The data that support the findings of this study are available from the corresponding author upon reasonable request.